\documentclass[11pt,a4paper]{article}
\usepackage[makeroom]{cancel}
\usepackage{bm}
\usepackage{amsmath}
\usepackage{amsfonts}
\usepackage{amssymb}
\usepackage{graphicx}
\usepackage{authblk}
\usepackage{cite}
\usepackage[left=2.4cm,top=2.4cm,right=2.4cm,bottom=2.4cm,nohead]{geometry}
\DeclareGraphicsExtensions{.png,.jpg,.pdf}
\usepackage{epstopdf}
\usepackage{float}
\usepackage{fouriernc} 
\usepackage{enumitem}
\usepackage[utf8]{inputenc}
\usepackage[english]{babel}
\usepackage{mathtools}
\usepackage{amscd}
\usepackage{mathrsfs}
\usepackage{amsthm}
\usepackage{siunitx}
\usepackage{caption}
\usepackage{subcaption}
\setlist{nolistsep,leftmargin=*}

\usepackage{mhchem} 
\usepackage{cleveref} 
\usepackage{xcolor}
\usepackage{soul}
\usepackage{mdframed}
\usepackage{framed}
\usepackage{longtable}
\usepackage{booktabs}
\usepackage{tabularx}
\DeclareMathAlphabet{\mathpzc}{OT1}{pzc}{m}{it}

\title{Analysis of the Rankine attraction term in an equation of state based on the London dispersion force}

\makeatletter

\renewcommand\AB@authnote[1]{\textsuperscript{\normalfont#1}}

\makeatother

\author{P.M. Biesheuvel}

\affil{Wetsus, European Centre of Excellence for Sustainable Water Technology,  
The~Netherlands.}

\date{}

\newcommand{\s}[1]{\mathrm{_{#1}}}

\begin{document}

\maketitle

\begin{abstract}

The attraction term in an equation of state for gases, $-a c^2$, proposed by Rankine in 1854, is generally related to the London dispersion force via the equation for the second virial coefficient, $B_2$, given by $B_2 = 2\pi N_0 \int_0^\infty \left(1- \exp \left(-\omega / kT\right)\right) r^2 \text{d}r$, where $\omega$ is the attraction energy between two molecules in the gas. This equation works very well to describe $B_2$ and thus $a=B_2 RT$ as function of temperature, but the derivation is complicated. Here we present two other methods to derive $B_2$ and thus $a$ from the London equation, which have a more intuitive background. (The more simple of the two we expect must be available in literature.) We analyze these three models for the gas argon at temperatures between 150 and 900 K. All three methods fit the data quite closely while an analytical approximation fits data even better. The temperature dependence of $a$ is well described by a proportionality of $a$ with a term $1+T^*/T$ where $T^*$ is a constant that depends on the type of gas. 

\end{abstract} 

In 1854, Rankine proposed to extend the ideal gas law with an attraction term, $-a c^2$, where \textit{c} is concentration, and \textit{a} an attraction factor. For atoms or molecules described as hard spheres, a correction due to volume exclusion was proposed by Jäger in 1896 as $P\s{hs}=c RT  \cdot \left(4 \eta + 10 \eta^2 + \dots \right)$, where $\eta$ is the volume fraction, $\eta = v\s{m} c$, with $v\s{m}$ the molecular volume (also used by Guggenheim in 1965). The complete equation of state then becomes 
\begin{equation}
P=P\s{id}+P\s{hs}+P\s{att}=c RT \cdot \left(1+ 4 \, \eta + 10  \, \eta^2 + \dots \right)  - a c^2 \, .
\end{equation}
In this paper we focus on the attraction term. Rankine assumed that the factor \textit{a} is inversely proportional to temperature, i.e., it depends on temperature according to $1/T$, i.e., it scales with temperature to the power $-1$. Later, Redlich and Kwong (1949) reduced the dependence of \textit{a} on temperature to a proportionality with $1/\sqrt{T}$, i.e., $a$ scales with temperature to the power ${-1/2}$. 

The question is whether the quadratic dependence of the attraction term can can be derived from (i.e., is predicted by) the London dispersion force, and what is then the dependence of \textit{a} on temperature. The London dispersion force describes that the energy of intermolecular attraction depends on one over the sixth power of distance between the molecules. In literature, the expression for the London dispersion force is then implemented in an equation for the second virial coefficient, $B_2$, given by $B_2=2 \pi N_0 \int \left(1- \exp \left(- \omega / kT\right)\right) r^2 \text{d}r$. The attraction part of the second virial coefficient, $B_\text{2,att}$, then relates to the attraction parameter \textit{a} according to $a = B_\text{2,att} RT$. In this approach, it is assumed that the attraction term has a quadratic dependence on \textit{c}. Assuming that $\omega$ is independent of temperature, then for a high enough temperature, $B_2$ has an inverse proportionality on \textit{T}, and thus the virial equation predicts that the attraction parameter, \textit{a}, is independent of temperature. But for lower temperatures, the predicted dependence of $B_2$ and \textit{a} on temperature can be different. 

\newpage

We will evaluate $B_2$ and $a$ of a gas due to the London dispersion force (attractive force) by the virial equation, as well as by two other methods. 
The advantage of the two alternative methods is that they have a more intuitive background than the virial equation which is based on a quite complicated derivation. The three methods give close to the same results, and give a close fit to data. This correspondence with data, and between the methods, is very important, because it gives support to allowing the use of one method instead of another for other systems of gas molecules or ions. 

Of the two alternative methods, we first describe the more detailed and complex method. We consider a gas molecule located at the center of a spherical coordinate system (at $r\!=\!0$), and we evaluate the distribution around it of the first, second, third, etc., closest neighbors, and calculate the energy of interaction of each of them with the molecule at the center. Then we multiply this result with concentration, to arrive at an energy density (also divide by 2), and take an appropriate derivative to obtain the contribution to pressure. 

The energy of attraction due to London dispersion is 
\begin{equation}
\omega=-\frac{\lambda}{r^6}
\label{eq_London}
\end{equation}
with $\lambda$ the London dispersion factor, also written as $C\s{disp}$, with unit J.m\textsuperscript{6}. It relates to the polarizability of the gas molecules, $\alpha$, and to an energy $h\nu$, according to $\lambda = \tfrac{3}{4} \alpha^2 h\nu$, where \textit{h} is the Planck constant and $\nu$ the orbiting frequency of an electron. Dependent on the molecule, this energy is in the range $h\nu=\left\{1.8-3.4\right\} \cdot 10^{-18}$~J. The polarizability of each molecule (atom) is tabulated, such as $\alpha=0.21~\AA^3$ for helium, and $\alpha=1.64~\AA^3$ for argon. The London dispersion factor, $\lambda$, is independent of temperature.\footnote{And for that reason, the Hamaker constant, $A=\pi^2 q^2 \lambda$, is temperature-independent as well; here, \textit{q} is the number density of atoms in a material.} In the derivation below, $\lambda$ is equated to the product of \textit{kT} and a specific length, $r\s{x}$, to the power 6, i.e., $\lambda=kT \cdot r_\text{x}^6$. Because $\lambda$ does not depend on temperature, $r_\text{x}^3$ is inversely proportional to the square root of temperature, $r_\text{x}^3\propto 1/\sqrt{T}$. 

The energy of attraction of a molecule with its nearest neighbor, due to London dispersion, is 
\begin{equation}
\Omega = \frac{\int \exp \left( -\omega/kT \right)  \cdot \omega \cdot N_1\, \text{d} r }{\int \exp \left( -\omega/kT \right) \hphantom{\cdot \omega}  \cdot N_1 \, \text{d} r }
\label{eq_Omega_nnd}
\end{equation}
where the integration limits go from two times the radius of the atoms, $r=2a$ (assuming the atoms are spherical), to infinity. We implement here a function $N_1$, which is the \textit{nearest neighbor distribution function}, given by the Hertz equation
\begin{equation}
N_1 = 4 \pi r^2 \cdot n \cdot \exp \left(- \tfrac{4}{3} \pi r^3 n \right) 
\label{eq_nnd}
\end{equation}
%
%
%
%
%
where \textit{n} is a concentration in numbers per volume, $n=c \, N_0$, with $N_0$ the Loschmidt number ($N_0=6.022\cdot 10^{23}$~mol\textsuperscript{-1}).

We can implement Eqs.~\eqref{eq_London} and~\eqref{eq_nnd} in Eq.~\eqref{eq_Omega_nnd}, and arrive at
\begin{equation}
\frac{\Omega}{kT} = -  \frac{\int \exp \left( {r_\text{x}^6}/{r^6} \right) \cdot {r_\text{x}^6} \cdot r^{-4} \cdot  \exp \left(-\tfrac{4}{3} \pi r^3 n \right) \, \text{d} r }{\int \exp \left( {r_\text{x}^6}/{r^6} \right) \hphantom{\cdot r_\text{x}^6} \cdot r^{2\hphantom{-}} \cdot \exp \left(- \tfrac{4}{3} \pi r^3 n  \right) \, \text{d} r } \, .
\label{eq_Omega_nnd_2}
\end{equation}
Next, we implement a dimensionless, inverse, coordinate, $y=r_\text{x}/r$, and then obtain 
\begin{equation}
\frac{\Omega}{kT} = - \frac{\int \exp \left( y^6 \right)  \cdot y^{2} \hphantom{^{-}} \cdot \exp \left(-\delta \cdot y^{-3}  \right) \, \text{d} y }{\int \exp \left( y^6 \right) \cdot y^{-4} \cdot \exp \left(- \delta \cdot y^{-3} \right) \, \text{d} y }
\label{eq_Omega_nnd_3}
\end{equation}
where $\delta = \tfrac{4}{3} \pi r_\text{x}^3 n$. Interestingly, the only point where concentration enters the calculation is via $\delta$. Temperature enters the calculation via $r_\text{x}$, so also via $\delta$, and in addition via the upper integration limit (which also depends on the atom size). 

Eqs.~\eqref{eq_Omega_nnd_2} and~~\eqref{eq_Omega_nnd_3} only consider interaction with the first neighbor, but we can simply add contributions of the second, third, etc., neighbor. We must only replace in each case the function for ${N}_1$ by the general expression for the distribution of the \textit{i}th neighbor, given by (Berberan Santos, 1987) 
\begin{equation}
N_i=\frac{3}{\left(i-1\right)!} \left( \tfrac{4}{3}\pi n \right)^i r^{3i-1} \exp \left(-\tfrac{4}{3} \pi r^3 n\right) \, .
\label{eq_ith_neighbor}
\end{equation}

The generalization of Eq.~\eqref{eq_Omega_nnd_3} then becomes
\begin{equation}
\frac{\Omega}{kT} = - {\sum}_i\left\{\frac{\int \exp \left( y^6 \right)  \cdot y^{\hphantom{-}5-3i} 
\cdot \exp \left(-\delta \cdot y^{-3}  \right) \, \text{d} y }{\int \exp \left( y^6 \right) \cdot y^{-1-3i} \cdot \exp \left(- \delta \cdot y^{-3} \right) \, \text{d} y } \right\} \, .
\label{eq_Omega_nnd_4}
\end{equation}

In a first calculation, we use $\alpha\!=\!1.63~\AA^3$ and $h \nu = 2.53 \cdot 10^{-18} $~J, which are reported values for argon, and thus the London dispersion constant is $\lambda=5.0\cdot 10^{-78}$~J.m\textsuperscript{6}. We calculate $r_\text{x}^3$ from $r_\text{x}^3=\sqrt{\lambda / kT}$, and thus at $T\!=\! 300$~K, we have $r_\text{x}\!=\!3.27~\AA$. We take a radius $a \! = \! 1.72~\AA$, and numerically integrate Eq.~\eqref{eq_Omega_nnd_4} from $y\!=\!0$ (infinitely far) to $y=r_\text{x}/\left(2a\right)$, which at 300 K is $y\!=\!0.950$. (A reported value for the kinetic radius of argon is 0.170 nm, thus we use a value very close to this.)

After generating calculation output for $\Omega$ as function of $\delta$, thus as function of concentration $n$, we find the pressure according to
\begin{equation}
P\s{att}= \tfrac{1}{2} \, n^2 \, \frac{d\Omega }{ dn }  \, .
\end{equation}
We make this calculation based on adding together the interaction with the first 5 neighbors (in this temperature range this is enough to have convergence), and evaluate this attractive pressure due to London dispersion, $P\s{att}$, in a range of concentrations, \textit{c}, from 0 to 500 mM, and find a perfect quadratic dependence, with a prefactor $a=-0.122$~J.m\textsuperscript{3}/mol\textsuperscript{2}. Dividing $a$ by \textit{RT} results in an attraction contribution to the second virial coefficient of $B\s{2,att} = - 49$~cm\textsuperscript{3}/mol. That is much more negative than an aggregate value for argon from literature that is $B_2=-16$~cm\textsuperscript{3}/mol, but in that number also the hard sphere effect is included. That contribution we estimate as $B\s{2,hs}=34$~cm\textsuperscript{3}/mol ($B_\text{2,hs}=4 v\s{m}$, with $v\s{m}$ a molecular volume), and then $B\s{2,att,lit}=-50$~cm\textsuperscript{3}/mol, which is close.

We redo the entire calculation at $T\!=\!400$~K and 500~K. From the reported aggregate values of $B_2$, we again subtract $B\s{2,hs}$, and thus assume the literature values for attraction are $B\s{2,att,lit} = -35$~and $-27$~cm\textsuperscript{3}/mol, respectively. We again find that the contribution of attraction to the pressure scales with concentration squared, and for $B\s{2,att}$ we calculate for 400 and 500 K values of $B\s{2,att}=-35$~and $-27$~cm\textsuperscript{3}/mol, respectively, in perfect agreement with the experimental data.  
Calculations at 600~K and 700~K result in $B\s{2,att}$ given by $-22$ and $-19$~cm\textsuperscript{3}/mol, resp., while the literature data for $B\s{2,att,lit}$ are exactly the same. 
At $T\!=\!200$~K, we calculate $B\s{2,att}=-85$~cm\textsuperscript{3}/mol, while $B\s{2,att,lit}=-82$~cm\textsuperscript{3}/mol, again close. At $T\!=\!160$~K, $B\s{2,att}=-118$~cm\textsuperscript{3}/mol, while $B\s{2,att,lit}=-110$~cm\textsuperscript{3}/mol, thus a deviation develops now that we come closer to the critical point. Interestingly, at these temperatures of 200 K and lower, the quadratic dependence is only closely followed at concentrations up to $\sim 200$~mM, and for the pressure curve to `stay quadratic' up to higher concentrations, we likely have to incorporate more and more neighbors than we did (we always analyzed only the first 5 neighbors). 
%

In an alternative method, we do not consider the first, second, third neighbors, etc., but assume a homogeneous distribution of molecules over space with a background concentration \textit{n}. The energy of interaction of the central molecule with all molecules around it, $\Psi$, is then given by
\begin{equation}
\frac{\Psi}{n} = 4 \pi \int \exp \left(-\omega/kT\right) \cdot \omega \cdot r^2 \, \text{d}r 
\end{equation}
where we again integrate from twice the molecular radius, to infinity. To calculate $B_\text{2,att}$, we multiply the right-hand side by $N_0^2$ and divide by $2RT$, which we can rewrite to
\begin{equation}
B_{2} = 2 \pi N_0 \int \exp \left(-\omega/kT\right) \cdot \left( \omega / kT \right) \cdot r^2 \, \text{d}r \, .
\label{eq_hom}
\end{equation}
This calculation gives results that are slightly different from the nearest neighbor-method, with $B_2$ slightly more negative for $ T < 300$~K, and slightly less negative for higher temperatures. If we go to temperature of 200 K and lower, it starts to deviate more significantly and gives too negative values of $B_\text{2,att}$. 

\begin{figure} \centering
\includegraphics[width=1.0\textwidth]{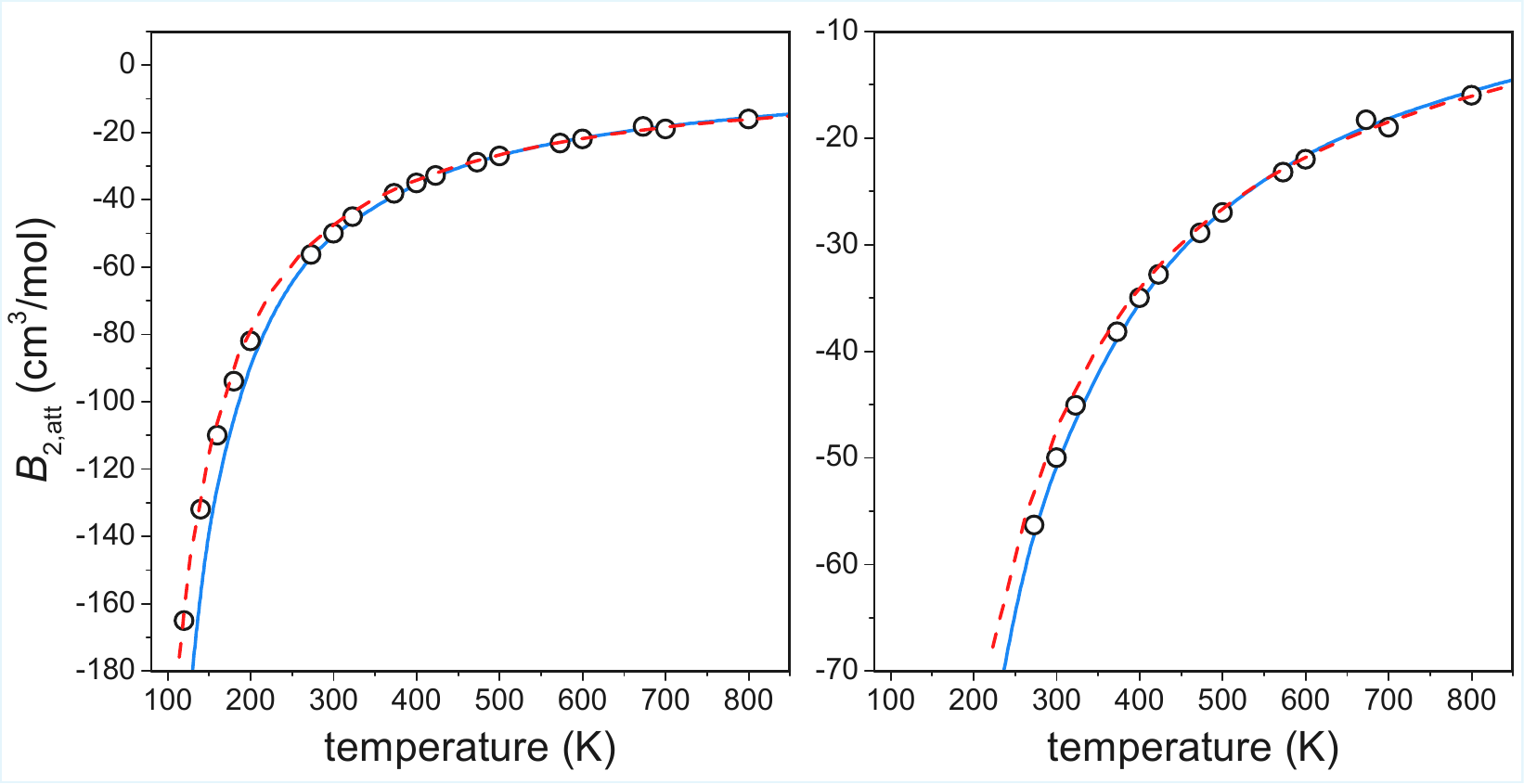}
\vspace{-25pt} 
\caption{The contribution of attraction to the second virial coefficient of argon, as function of temperature. The red dashed line is according to the virial route, and continuous blue line is based on evaluation of the London dispersion force by the homogeneous method. 
Circles are data from literature of aggregate $B_2$-values from which a hard sphere-contribution, $B_\text{2,hs}=34$~cm\textsuperscript{3}/mol, is subtracted.}
\label{fig_B2_argon}   
\end{figure}

We also solve $B_\text{2,att}$ according to the virial approach, $B_2=2 \pi N_0 \int \left(1- \exp \left(-\omega / kT\right)\right) r^2 \text{d}r$, using the same $\lambda$, but a slightly lower molecular radius of $a=0.168$~nm to fit data as good as possible. Data are described very well, only overestimating the data of $B_\text{2,att}$ slightly for temperatures below 400 K, up to about 5 point cm\textsuperscript{3}/mol at 120~K. 
For the homogeneous method and the virial route, the calculated values for $B_\text{2,att}$ are plotted in Fig.~\ref{fig_B2_argon} as a continuous blue line and a red dashed line, respectively, together with literature data. In the full range, the virial route is better, but considering only data above 200~K, the situation is reversed.

\begin{figure} \centering
\includegraphics[width=0.6\textwidth]{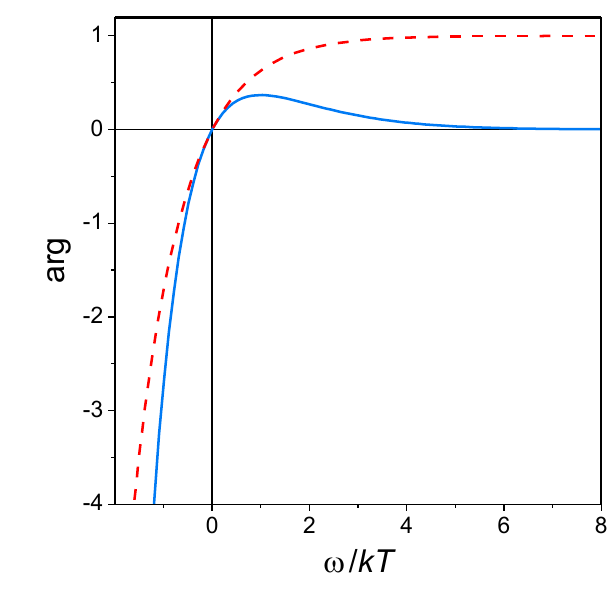}
\vspace{-10pt} \caption{Comparison of the virial route (dashed red line) with the homogeneous method, Eq.~\eqref{eq_hom} (continuous blue line), plotting the argument in the integration as function of $\omega / kT$ (ignoring $r^2$). The two methods overlap for small $\omega / kT$, but they are different for lower and for higher energies.}
\label{fig_hom_vs_vir}   
\end{figure}

We compare in Fig.~\ref{fig_hom_vs_vir} the argument in the integration of the homogeneous method, Eq.~\eqref{eq_hom}, with that of the virial route, $1-\exp\left(\omega / kT\right)$. For low $\omega / kT$, the two methods overlap, but both at lower and higher energies they deviate from one another. At more negative $\omega / kT$, the homogeneous method gives a more negative prediction of $B_2$ than the virial route, while for more positive $\omega / kT$, the virial route has an argument that approaches 1, but for the homogeneous route the argument has a maximum and then goes to zero. Because of this difference, the virial route can also correctly predict the hard sphere-repulsion, $B_\text{2,hs}=4 v\s{m}$, but the homogeneous method cannot.

We can make a series expansion of the homogeneous method and the virial route, around $\omega=0$. For low $\omega/kT$, the two methods are the same (the first term is $\omega/kT$), but the second term is $-\omega^2/\left(kT\right)^2$ for the homogeneous route, and half of that for the virial route. After integration of these first two terms for the homogeneous route, we obtain the analytical solution
\begin{equation}
B_\text{2,att}= - 12 \cdot v_\text{m}^* \cdot T^* /  T \cdot \left( 1 + T^* / T\right)
\end{equation}
where $v_\text{m}^*=\frac{4}{3} \pi a^3 N_0$ is a molecular volume based on the radius $a$, and $T^* = \lambda / \left( 192 \cdot a^6 \cdot k \right)$ is a temperature factor (unit K). When we use $a=0.170$~nm, and thus $v_\text{m}^*=12.4$~cm\textsuperscript{3}/mol, this function fits the data for argon exactly at all temperatures, better than all methods discussed up to now. 

Thus, we propose for argon in the temperature range $150-900$~K the equation of state
\begin{equation}
P = c RT \cdot \left(1 + 4 \, \eta +10 \,\eta^2 + 20 \, \eta^3\right) - 12 \cdot R \cdot v_\text{m}^* \cdot T^* \cdot   \left( 1 + T^* / T\right) \cdot c^2
\label{eq_EOS_ar}
\end{equation}
where $\eta = v\s{m} c$ is based on $v\s{m}=8.5$~cm\textsuperscript{3}/mol, while for argon, $T^*=79.6$~K. Here we also added the next two terms in the Jäger expression for the hard sphere repulsion (nowadays, the factor 20 is replaced by 18 in the Carnahan-Starling equation of state). With these numbers, argon is athermal at $T=415$~K, i.e., in the dilute limit, volume effects and attraction cancel each other out at that temperature, and it then behaves as an ideal gas. Based on Eq.~\eqref{eq_EOS_ar}, the critical point is when $dP/dc=0$ and $d^2P/dc^2=0$, which results in $P\s{c}=8.9$~MPa, $T\s{c}=187$~K, and $c\s{c}=16.4$~M. Experimental data are $P\s{c,exp}=4.9$~MPa, $T\s{c,exp}=150.8$~K, and $c\s{c,exp}=13.4$~M, and thus this simple theory that was only evaluated at low concentrations ($< 0.5$~M) is definitely off, but not dramatically. 


In the Redlich-Kwong theory, the attraction term is modified by dividing by $1+bc$, where \mbox{$b=4 v\s{m}$}. We implement this term in Eq.~\eqref{eq_EOS_ar}, resulting in
\begin{equation}
P = c RT \cdot \left(1 + 4 \,\eta +10 \,\eta^2 + 20 \, \eta^3\right) - 12 \cdot R \cdot v_\text{m}^* \cdot T^* \cdot   \left( 1 + T^* / T \right) \cdot c^2 { / \vphantom{\tfrac{1}{8}}  }  \left( 1 + \alpha \eta \right)
\label{eq_EOS_ar_2}
\end{equation}
which gives an improved description of the critical point, especially when we allow the factor $\alpha$ not to be 4, but to be reduced to $\alpha \! = \! 1.9$. Then the equation predicts $P\s{c}=4.9$~MPa, $T\s{c}=149$~K, and $c\s{c}=12.0$~M, which are predictions close to the data. 
Thus, small modifications to the attraction term, that at low concentrations do not have an effect, can markedly modify the prediction of the critical point. In this case, the attraction term is modified by $\sim 15\%$ at the critical point when we include $1+\alpha \eta$. The temperature factor $T^*$ depends on the type of gas via the kinetic radius, $a$, and London coefficient, $\lambda$. We calculate for helium, neon, argon, and xenon, the following temperature factors: $T^* = 15.7,~22.0,~79.6,~\text{and}~128.3$~K, resp.

In conclusion, we presented an analysis of the London dispersion force based on the virial route commonly used in literature, and two alternative methods, one where a homogeneous concentration is considered, and one including in more detail interactions of a molecule with its first, second, third, etc., neighbors. For argon in the temperature range $150-900$~K, we find that all methods are close to the data, while an analytical equation fits data even better. In this analytical equation, the attraction parameter, \textit{a}, in the Rankine attraction term, $-a c^2$, depends on temperature according to $a \propto 1+T^*/T$ with \textit{T} a material-specific temperature factor. This dependence implies that the scaling of \textit{a} with temperature is for instance $a \propto T^{-1/10}$ at a temperature of 700~K, and $a \propto T^{-1/5}$ at 300~K. A small volume correction to the attraction term due to Redlich-Kwong has a marked effect on the prediction of the critical point of argon.

\newpage


\end{document}